**Microwave-Induced Cooling in Double Quantum Dots: Achieving milliKelvin Temperatures to Reduce Thermal Noise around Spin Qubits**

Daryoosh Vashaee[a,b*] and Jahanfar Abouie[c*]
[a] Department of Materials Science & Engineering, North Carolina State University, Raleigh, NC 27606, USA
[b] Department of Electrical & Computer Engineering, North Carolina State University, Raleigh, NC 27606, USA
[c] Department of Physics, Institute for Advanced Studies in Basic Sciences, Zanjan 45137-66731, Iran

**Abstract**

Spin qubits in gate-defined quantum dots (QDs) are emerging as a leading technology due to their scalability and long coherence times. However, maintaining these qubits at ultra-low temperatures typically requires complex cryogenic systems. This paper proposes a novel gate-defined double quantum dot (DQD) cooling system, where the DQDs act as refrigerants to reduce the local phonon environment around computational qubits. The cooling process occurs in two distinct stages: the first step involves microwave-induced state depopulation combined with fast cyclic detuning to transfer the DQD's population to the ground state, effectively lowering the DQD's temperature. In the second step, the cooled DQD interacts with and absorbs phonons resonant with the DQD spin energy, thereby filtering out these phonons that contribute to spin-lattice relaxation in the surrounding environment. This study focuses on the first step, presenting detailed calculations and numerical results that demonstrate the feasibility of achieving local DQD temperatures below $10\ \mathrm{mK}$ at a bath temperature of $1\ \mathrm{K}$. The sensitivity of the cooling performance to detuning energy, magnetic field strength, and diabatic return time is analyzed, while the phonon filtering in the second step will require further investigation.

**Introduction**

Quantum computing promises to revolutionize various fields by solving problems that are intractable for classical computers. Among the different qubit technologies, spin qubits in gate-defined quantum dots (QDs) have emerged as a leading candidate due to their scalability and relatively long coherence times.[1] These qubits typically operate at sub-100 $\mathrm{mK}$ temperatures, which presents a significant challenge for large-scale integration of control electronics and other components.[2] There is a substantial technological interest in increasing their operational temperature to above $1\ \mathrm{K}$, which would facilitate easier integration and reduce the complexity of cooling systems.[3] Achieving this would not only simplify the infrastructure but also enhance the practicality and scalability of quantum computing systems.

Cooling is a critical aspect of quantum computing, particularly for maintaining qubit coherence and minimizing thermal noise. For gate-defined QD spin qubits, operating at higher temperatures while maintaining high fidelity is a significant challenge. Traditional methods such as dilution refrigerators are effective but cumbersome and expensive. Innovations in cooling technologies can thus have profound implications for the development and deployment of quantum computers.

Over the past several decades, various cooling technologies have been developed for different qubit systems. Sympathetic cooling for ion traps involves cooling one species of ions using laser cooling and then transferring the cooling effect to another species through Coulomb interactions.[4] This technique has been extended to spatially separated traps using superconducting LC circuits, enabling efficient cooling over macroscopic distances.[5] Sideband cooling for superconducting qubits has been demonstrated, where thermal populations are driven to higher excited states via sideband transitions, followed by relaxation to the ground state. This method has achieved effective temperatures as low as $3\ \mathrm{mK}$ for bath temperatures ranging from $30$ to $400\ \mathrm{mK}$.[6] Algorithmic cooling leverages quantum algorithms to transfer entropy from some qubits to others or to the environment, effectively cooling the system. Algorithmic cooling has been shown to exceed Shannon's bound on data compression and has applications in initializing highly pure qubits for quantum computation.[7,8]

* Corresponding authors: dvashae@ncsu.edu and jahan@iasbs.ac.ir

Recent advancements propose the use of few-qubit quantum refrigerators for cooling multi-qubit systems. These systems typically involve a central qubit coupled to several ancilla qubits, using short interactions to achieve cooling.[9,10,11] However, these methods are usually applicable for the initialization stage and do not work for computational qubits during operations. [9,12] Additionally, the number of ancilla qubits required to achieve sub-100 $\mathrm{mK}$ temperatures increases rapidly, leading to greater size and cost for quantum technology applications.[9,13] Therefore, there remains a gap in effectively cooling interacting multi-qubit systems, particularly those based on spin qubits in quantum dots. This work aims to address this gap by exploring a DQD system as an alternative approach for quantum refrigeration. This new method could provide a more efficient and practical solution for cooling at higher operational temperatures using a single DQD, thereby facilitating the integration of control electronics and enhancing the scalability of quantum computing systems.

The structure of this paper is as follows. In **Sec. I**, we introduce our cooling mechanism for the double quantum dot (DQD) system. **Sec. II** presents the theoretical formalism, detailing the Hamiltonian of the DQD and evaluating its eigenenergies and eigenvectors through analytical diagonalization. This section also briefly addresses the adiabatic condition for an N-level system. In **Sec. III**, we discuss the thermal and microwave excitations and describe the cooling cycle process. This section includes our primary numerical results and explains energy conservation during gate voltage transitions. **Sec. IV** outlines the necessity of a DQD system for effective cooling, comparing it with the limitations of a single QD under microwave excitation. **Sec. V** addresses practical considerations, and **Sec. VI** summarizes our findings.

## I. Microwave-Induced Cooling Mechanism for Double Quantum Dots

We propose leveraging the interplay among state transitions of a double quantum dot, microwave absorption at resonance, and detuning of the potential of the QDs to design a cooling cycle capable of reducing the system temperature to the milliKelvin (mK) range using hole or electron spins. This cycle exploits the rapid response of hole spin qubits via spin-orbit coupling (SOC) to implement a dual cooling mechanism involving (i) microwave-induced state depopulation and (ii) phonon filtering (Figure 1). By adjusting gate potentials, the system is manipulated to drive the thermal population back to the ground state, surpassing natural thermal transition rates. Additionally, this method uses the energy imbalance induced by microwave absorption to extract energy from specific phonon modes, significantly lowering the system's energy. Repeating this process cyclically can achieve temperatures well below the surrounding environment.

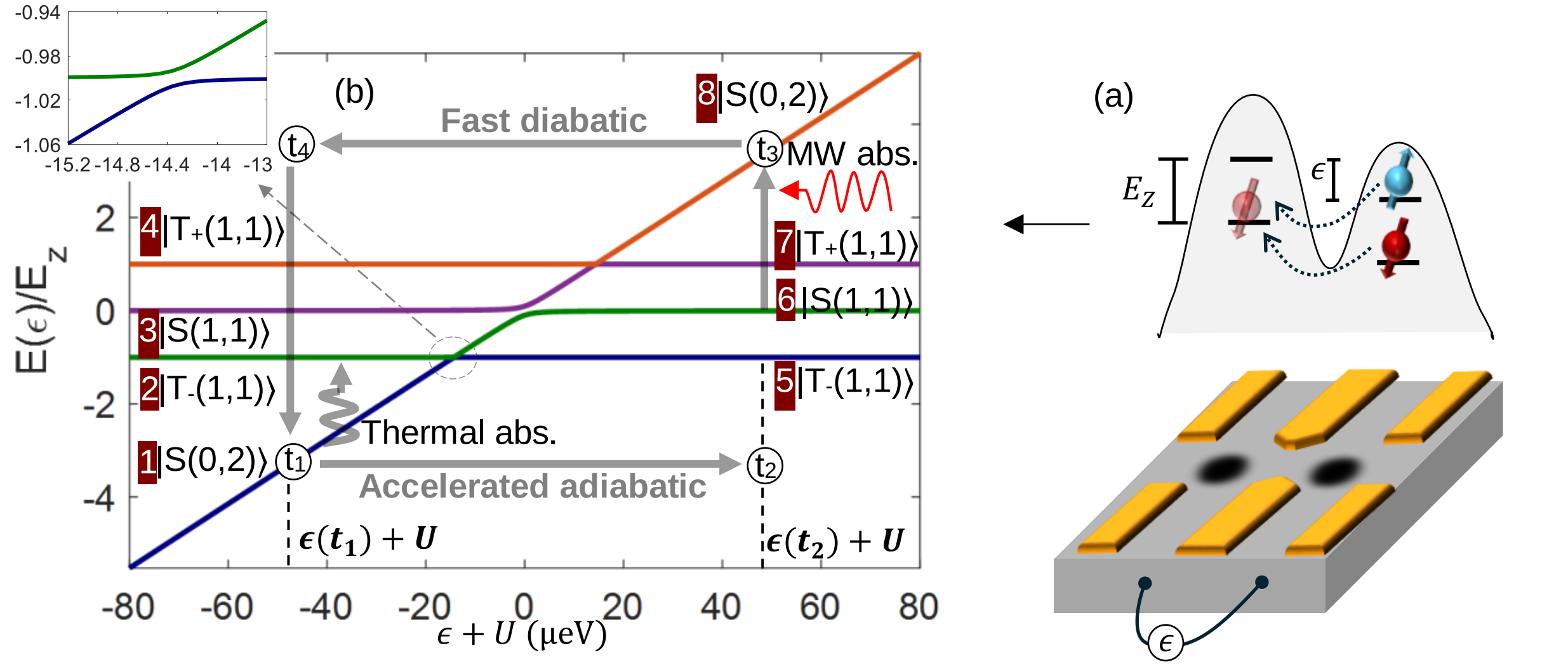

**Figure 1:** (a) A gate-defied DQD with two HH spins with Zeeman splitting $E_Z$ and level detuning $\epsilon = \epsilon_R - \epsilon_L$. Both the spin-conserving and spin-flipping tunneling are allowed through DQD in the presence of SOC (see the section of theoretical formalism). The level detuning energy is controlled by the gate potential. (b)The lowest energy levels of the Hamiltonian $H$ in the plotted energy range. Here, $\epsilon(t_1)$ and $\epsilon(t_2)$ are the difference between the on-site energies at times $t_1$ and $t_2$, respectively. Initially, thermal excitations transfer population fractions from the spin ground state $|S(0,2)\rangle$ to the excited states $|T_-(1,1)\rangle$, $|S(1,1)\rangle$, and $|T_+(1,1)\rangle$ according to $p_i = \exp(-E_i/k_B T_0)/\sum_i \exp(-E_i/k_B T_0)$, where $p_j$ is the population fraction at energy level $E_j$, and $T_0$ is the bath temperature. To cool below $T_0$, energy is cycled between two adjacent QDs by detuning from $\epsilon(t_1)$ to $\epsilon(t_2)$ using an accelerated adiabatic passage, followed by microwave absorption and then quickly returning to $\epsilon(t_1)$ via a fast diabatic path. This process transfers the excited population to the ground state, cooling the DQD. The microwave signal applied to the DQD at time $t_2$ induces a transition from $|S(1,1)\rangle$ to $|S(0,2)\rangle$ . Upon quick returning to $\epsilon(t_1)$, the system remains at $|S(0,2)\rangle$but at a lower energy, effectively driving the thermal population back to the ground state. This transition must occur faster than the tunneling rate to prevent charge transfer to $|T_\pm(1,1)\rangle$ or $|S(1,1)\rangle$ .

Initially, the system is at a detuning energy $\epsilon(t)$. The ground state at this point is $|S(0,2)\rangle$, with the first excited state $|T_-(1,1)\rangle$, the second excited state $|S(1,1)\rangle$, and the third excited state at $|T_+(1,1)\rangle$ (labeled as 1, 2, 3, and 4, respectively, in Figure 1). The system transitions at time $t_1$ from $\epsilon(t_1)$ to $\epsilon(t_2)$ by tuning the differential potential across the gates. This transition is executed through accelerated adiabatic transitions over time $t_{12} = t_2 - t_1$. Consequently, populations $p_1, p_2, p_3$, and $p_4$ transition to $p_5, p_6, p_7$, and $p_8$, respectively. At time $t_2$, a microwave pulse resonant with the energy difference between states $|6\rangle$ and $|8\rangle$ transitions the $p_6$ population to $p_8$. The pulse duration is adjusted so that the entire $p_6$ (and $p_5$ and $p_7$) population is transferred to $p_8$ at time $t_3$.

At $t_3$, the potential across the gates is quickly returned to the starting voltage, setting $\epsilon(t_4) = \epsilon(t_1)$. This transition must be faster than the tunneling rate to ensure the charge remain in the $|S(0,2)\rangle$ configuration. Without charge transfer, there is no population transfer to $p_2, p_3$, and $p_4$. Consequently, the entire population is transferred to the ground state $|1\rangle$. This process moves the population in the excited states $|2\rangle$, $|3\rangle$, and $|4\rangle$ to the ground state $|1\rangle$, leading to cooling of the DQD system.

The second aspect, microwave-induced phonon filtering, happens at the end of the first cycle. At this time, the DQD is at a lower local temperature compared to the lattice, causing it to absorb phonons to reach equilibrium with the environment. By cycling the DQD back and forth between the two detuning points, more phonons are absorbed, effectively filtering the phonons locally where the DQD is situated. Phonons resonant with the energy difference of states $|1\rangle$ and $|2\rangle$ are expected to be filtered more efficiently. This energy can be adjusted to match the Zeeman energy splitting of a nearby computational spin qubit. Therefore, phonons in resonance with the Zeeman energy splitting of the computational qubit can be filtered around the qubit. Consequently, the cold DQD acts as a filter for specific phonon modes, particularly those resonant with the energy levels of the spin qubit. By selectively interacting with these local phonon modes, the cold DQD reduces the effective phonon noise experienced by the spin qubit, enhancing its coherence time.

The following sections describe the theory and computational results of the proposed DQD cooler.

**II. Theoretical Formalism**

We examine a planar DQD system containing two heavy holes (HHs), as depicted in Figure 1(a). The overall Hamiltonian is expressed as $H = H_{DQD} + H_B + H_{SOC}$ where the first term is defined as:

$$H_{DQD} = \sum_{i=L,R} \sum_{\sigma=\uparrow,\downarrow} \epsilon_i c_{i\sigma}^\dagger c_{i\sigma} + U \sum_{i=L,R} n_{i\uparrow} n_{i\downarrow} - t_N \sum_{\sigma=\uparrow,\downarrow} \left( c_{L\sigma}^\dagger c_{R\sigma} + Hc \right) \quad (1)$$

Here, $\epsilon_i$ (for $i = L, R$) represents the energy levels of the left and right dots, $U$ is the intradot Coulomb energy (on-site Hubbard interaction), and $t_N$ is the spin-conserving tunneling amplitude between adjacent dots. The operators $c_{i\sigma}(c_{i\sigma}^\dagger)$ annihilate (create) a hole with spin $\sigma$ in dot $i$, and $n_{i\sigma}$ is holes number operator. A magnetic field $B$, perpendicular to the plane of the dots, induces a Zeeman splitting. The interaction of the magnetic field with the HHs in the DQD is given by the following Hamiltonian:

$$H_B = \frac{1}{2} g^* \mu_B B \sum_{i=L,R} (n_{i\uparrow} - n_{i\downarrow}) \quad (2)$$

where $g^*$ is the effective Landé factor and $\mu_\mathrm{B}$ is the Bohr magneton. We assume a homogeneous magnetic field and *g* factor. The spin-orbit coupling (SOC) is described by:

$$H_{SOC} = i\alpha E_\perp (\sigma_+ p_3^- - \sigma_- p_3^+) - \beta(\sigma_+ p_- p_+ p_- + \sigma_- p_+ p_- p_+) \quad (3)$$

The ladder operators are defined as $\sigma_\pm = \left(\sigma_x \pm i\sigma_y\right)/\sqrt{2}$, and the momentum operator $p_\pm = p_x \pm i p_y$ is given by $\boldsymbol{p} = -i\hbar\nabla + \frac{e^*}{c}\boldsymbol{A}$**,** where $e^*$ is the effective charge of the hole, $c$ is the speed of light, and $\boldsymbol{A}$ is the magnetic vector potential. This equation represents the Rashba SOC ($\alpha$) due to structure inversion asymmetry, controlled by the effective electric field $E_\perp$ from the accumulation gate, and Dresselhaus SOC ($\beta$) due to bulk inversion asymmetry.

Since we consider a closed system, it is convenient to write the Hamiltonian matrix in the molecular basis of singlets: $|S(1,1)\rangle \equiv (|\uparrow,\downarrow\rangle - |\downarrow,\uparrow\rangle)/\sqrt{2}$, $|S(0,2)\rangle \equiv |0,\uparrow\downarrow\rangle$ and $|S(2,0)\rangle \equiv |\uparrow\downarrow,0\rangle$, and triplets $|T_0(1,1)\rangle \equiv (|\uparrow,\downarrow\rangle + |\downarrow,\uparrow\rangle)/\sqrt{2}$, $|T_-(1,1)\rangle \equiv |\downarrow,\downarrow\rangle$, and $|T_+(1,1)\rangle \equiv |\uparrow,\uparrow\rangle$. The nonvanishing spin-flip tunneling matrix elements are between the polarized triplet states and the double- and single-occupied singlet states as:

$$\langle T_\pm(1,1) | H_{SOC} | S(0,2) \rangle = \lambda_2 \quad (4)$$

$$\langle T_\pm(1,1) | H_{SOC} | S(2,0) \rangle = \lambda_2 \quad (5)$$

$$\langle T_\pm(1,1) | H_{SOC} | S(1,1) \rangle = \lambda_1 \quad (6)$$

where $\lambda_1 = \sqrt{2} S \lambda_2$. The overlap between the wave functions in each dot is defined as $S = \langle L|R \rangle = \exp(-d^2/2l^2)$, where $d$ is the distance between dots and $l$ is the extent of the wave function centered at each dot. The matrix element $\lambda_2$ also depends on these parameters.

Under a constant magnetic field $B$, the unpolarized triplet state $|T_0(1,1)\rangle$ does not interact with any other states. The anticrossing between $|T_+(1,1)\rangle$ and the singlet states occurs at a detuning near $\epsilon + U \approx -2t_N^2/E_Z + E_Z$, where $\epsilon$ is the detuning energy controlled by the gate potential. Additionally, if the detuning is sufficiently large ($\epsilon > -U$), the double-occupied singlet state $|S(2,0)\rangle$ is energetically distant and does not play a role.

To obtain the eigenenergies and eigenvectors of the total Hamiltonian $H$, we construct its matrix in the basis $|T_-(1,1)\rangle, |S(0,2)\rangle, |S(1,1)\rangle, |S(2,0)\rangle, |T_+(1,1)\rangle, |T_0(1,1)\rangle$. Considering that $\epsilon = \epsilon_R - \epsilon_L$ and $\epsilon_R + \epsilon_L = 0$, the action of $H = H_{DQD} + H_B + H_{SOC}$ on these states are given by:

$$H|T_0(1,1)\rangle = H_{SOC}|T_0(1,1)\rangle = 0, \quad (7)$$

$$H|T_+(1,1)\rangle = E_z|T_+(1,1)\rangle + \lambda_2^*(|S(0,2)\rangle + |S(2,0)\rangle) + \lambda_1^*|S(1,1)\rangle, \quad (8)$$

$$H|T_-(1,1)\rangle = -E_z|T_-(1,1)\rangle + \lambda_2^*(|S(0,2)\rangle + |S(2,0)\rangle) + \lambda_1^*|S(1,1)\rangle, \quad (9)$$

$$H|S(0,2)\rangle = (\epsilon + U)|S(0,2)\rangle - \sqrt{2}t_N|S(1,1)\rangle + \lambda_2(|T_+(1,1)\rangle + |T_-(1,1)\rangle), \quad (10)$$

$$H|S(2,0)\rangle = (-\epsilon + U)|S(2,0)\rangle - \sqrt{2}t_N|S(1,1)\rangle + \lambda_2(|T_+(1,1)\rangle + |T_-(1,1)\rangle), \quad (11)$$

$$H|S(1,1)\rangle = -\sqrt{2}t_N(|S(0,2)\rangle + |S(2,0)\rangle) + \lambda_1(|T_+(1,1)\rangle + |T_-(1,1)\rangle), \quad (12)$$

where we have used:

$$\hat{H}_{soc}|T_+(1,1)\rangle = \lambda_2^*(|S(0,2)\rangle + |S(2,0)\rangle) + \lambda_1^*|S(1,1)\rangle, \quad (13)$$

$$\hat{H}_{soc}|T_-(1,1)\rangle = \lambda_2^*(|S(0,2)\rangle + |S(2,0)\rangle) + \lambda_1^*|S(1,1)\rangle, \quad (14)$$

$$\hat{H}_{soc}|S(0,2)\rangle = \lambda_2(|T_+(1,1)\rangle + |T_-(1,1)\rangle), \quad (15)$$

$$\hat{H}_{soc}|S(2,0)\rangle = \lambda_2(|T_+(1,1)\rangle + |T_-(1,1)\rangle), \quad (16)$$

$$\hat{H}_{soc}|S(1,1)\rangle = \lambda_1(|T_+(1,1)\rangle + |T_-(1,1)\rangle), \quad (17)$$

The vanishing act of the Hamiltonain on the unpolarized state indicates that this state does not interact with the other states. Thus, the total Hamiltonian $H$ is a $5 \times 5$ matrix, and is expressed in the basis $(|T_-(1,1)\rangle, |S(0,2)\rangle, |S(1,1)\rangle, |S(2,0)\rangle, |T_+(1,1)\rangle)$ as:

$$H = \begin{bmatrix} -E_z & \lambda_2 & \lambda_1 & \lambda_2 & 0 \\ \lambda_2^* & \epsilon + U & -\sqrt{2}t_N & 0 & \lambda_2^* \\ \lambda_1^* & -\sqrt{2}t_N & 0 & -\sqrt{2}t_N & \lambda_1^* \\ \lambda_2^* & 0 & -\sqrt{2}t_N & -\epsilon + U & \lambda_2^* \\ 0 & \lambda_2 & \lambda_1 & \lambda_2 & E_z \end{bmatrix} \quad (18)$$

Here, the Zeeman splitting in each quantum dot is given by $E_Z = g^*\mu_B B$, where $g^*$ is the effective $g$-factor, $\mu_B$ is the Bohr magneton, and B is the magnetic field. The terms $\lambda_1^*$ and $\lambda_2^*$ are the complex conjugates of $\lambda_1$ and $\lambda_2$, respectively. The spin-orbit coupling terms represented by $\lambda_1$ and $\lambda_2$ capture the spin-flip tunneling processes between the states. The diagonal terms account for the energy contributions from the Zeeman effect and the detuning energies, while the off-diagonal elements represent tunneling and SOCs.

For sufficiently small values of SOCs $\lambda_1$ and $\lambda_2$, as indicated by the relations in equation (1), the states $|T_\pm(1,1)\rangle$ are the eigenstates of the Hamiltonian H, with the eigenenergies $\pm E_z$. These eigenenergies are depicted in Figure 1 with the horizontal orange-purple and green-blue lines at $E(\epsilon)/E_z = \pm 1$. Moreover, for very small values of spin-conserved tunneling energy, the singlet states are also eigenstates of the system: the doubly occupied singlet states $|S(0,2)\rangle$ and $|S(2,0)\rangle$ with eigenenergies $\epsilon + U$ and $-\epsilon + U$, respectively, and the state $|S(1,1)\rangle$ with a vanishing eigenenergy. In Figure 1, the eigenstate $|S(0,2)\rangle$ is represented by the off-diagonal blue-green-purple-orange line with $E(\epsilon)/E_z = (\epsilon + U)/E_z$, and the eigenstate $|S(1,1)\rangle$ is represented by the purple-green horizontal line at $E_z = 0$. The eigenenergy $-\epsilon + U$ is a diagonal line that is distant from the other eigenenergies and does not appear in our working energy window.

By turning the SOCs and spin-conserved tunneling energy, states repulsions make deformations in the energy spectrum. These deformations are seen as three avoided level crossing points, one with $\lambda_2$-dependent energy gap appeared at $\epsilon + U = 0$, and two with $t_N$-dependent energy gaps appeared around the $\epsilon + U = 0$ point, symmetrically.

In the absence of the SOCs, the total Hamiltonian is block diagonal and the diagonalization of the following Hamiltonian

$$H_{3\times 3} = \begin{bmatrix} \epsilon + U & -\sqrt{2}t_N & 0 \\ -\sqrt{2}t_N & 0 & -\sqrt{2}t_N \\ 0 & -\sqrt{2}t_N & -\epsilon + U \end{bmatrix}, \quad (19)$$

gives the eigenstates of the system. The eigenenergies are obtained as:

$$E_0 = \frac{2U}{3} - \frac{U^2+3\epsilon^2+12t_N{}^2}{3\left(x_1+\frac{\sqrt{x_2}}{2}\right)^{\frac{1}{3}}} - \frac{1}{3}\left(x_1 + \frac{\sqrt{x_2}}{2}\right)^{\frac{1}{3}}, \tag{20}$$

$$E_\pm = \frac{2U}{3} - \frac{U^2+3\epsilon^2+12t_N{}^2}{\frac{3}{2}(-1\pm\sqrt{3}i)\left(x_1+\frac{\sqrt{x_2}}{2}\right)^{\frac{1}{3}}} - \frac{-1\pm\sqrt{3}i}{6}\left(x_1 + \frac{\sqrt{x_2}}{2}\right)^{\frac{1}{3}}, \tag{21}$$

with $E_0 < E_- < E_+$. The corresponding eignestates are given in terms of the singlet states as:

$$|\Phi_0\rangle = \frac{1}{\sqrt{\left(-1+\left(\frac{E_0}{\sqrt{2}t_N}\right)^2+\frac{E_0(\epsilon-U)}{2t_N{}^2}\right)^2+\left(\frac{E_0+\epsilon-U}{\sqrt{2}t_N}\right)^2+1}}\left(\left(-1+\left(\frac{E_0}{\sqrt{2}t_N}\right)^2+\frac{E_0(\epsilon-U)}{2t_N{}^2}\right)|S(0,2)\rangle - \left(\frac{E_0+\epsilon-U}{\sqrt{2}t_N}\right)|S(1,1)\rangle + |S(2,0)\rangle\right),$$

$$|\Phi_\pm\rangle = \frac{1}{\sqrt{\left(-1+\left(\frac{E_\pm}{\sqrt{2}t_N}\right)^2+\frac{E_\pm(\epsilon-U)}{2t_N{}^2}\right)^2+\left(\frac{E_\pm+\epsilon-U}{\sqrt{2}t_N}\right)^2+1}}\left(\left(-1+\left(\frac{E_\pm}{\sqrt{2}t_N}\right)^2+\frac{E_\pm(\epsilon-U)}{2t_N{}^2}\right)|S(0,2)\rangle + \left(\frac{-E_\pm-\epsilon+U}{\sqrt{2}t_N}\right)|S(1,1)\rangle + |S(2,0)\rangle\right), \tag{22}$$

Here, $x_1 = U(U^2 + 18t_N{}^2 - 9\epsilon^2)$, and $x_2 = -4(U^2 + 3\epsilon^2 + 12t_N{}^2)^3 + 4x_1{}^2$.

To achieve an adiabatic transfer from the triplet state $|T_-(1,1)\rangle$ to the singlet state $|S(1,1)\rangle$ in a DQD system, the procedure involves designing a detuning pulse that allows the system to evolve along the instantaneous eigenstate. This process is detailed in reference [14]. The goal is to transition from $|T_-(1,1)\rangle$ to $|S(1,1)\rangle$ by manipulating the detuning $\epsilon(t)$ between the quantum dots. The accelerated quasi-adiabatic protocol has been proposed to achieve high-fidelity state transfer [14]. The protocol adjusts the detuning $\epsilon(t)$ dynamically to maintain adiabatic conditions, minimizing diabatic transitions. In brief, the adiabatic condition for an $N$-level system is:

$$C = \hbar\sum_{k\neq i}^{N}\left|\frac{\langle\phi_i(t)|\partial_t H(t)|\phi_k(t)\rangle}{[E_i(t)-E_k(t)]^2}\right|, \tag{23}$$

where $|\phi_k(t)\rangle$ are the instantaneous eigenstates and $E_k(t)$ the corresponding eigenenergies, and $C$ is a dimensionless constant. By solving the differential equation:

$$\dot{\epsilon} = \frac{C}{\hbar}\sum_{k\neq i}^{N}\left|\frac{[E_i(\epsilon)-E_k(\epsilon)]^2}{\langle\phi_i(\epsilon)|\partial_\epsilon H(t)|\phi_k(\epsilon)\rangle}\right|, \tag{24}$$

the time-dependent driving parameter $\epsilon(t)$ is obtained, ensuring a constant value of $C$ during the transfer. The fidelity of the protocol is defined as $\mathcal{F} \equiv \left|\langle S(1,1)|\Psi(t_f)\rangle\right|^2$, where $|\Psi(t_f)\rangle$ is the finial state achieved subsequent of the accelerated adiabatic transfer. The protocol shows an ondulatory behavior of fidelity, tending asymptotically to unity as the total time $t_f$ increases. The boundary conditions for the detuning $\epsilon(t_1)$ and $\epsilon(t_2)$ significantly affect the transfer fidelity. Higher fidelities are achieved by increasing the detuning range $\Delta\epsilon = \epsilon(t_2) - \epsilon(t_1)$ [14].

### III. Thermal and Microwave Excitations

The four states at $\epsilon(t_1) + U$ are in constant interaction with phonons, leading to thermodynamic equilibrium at the bath temperature $T_0$. The population fraction rate equations in matrix form can be expressed as:

$$\frac{d}{dt}\begin{bmatrix}p_1\\p_2\\p_3\\p_4\end{bmatrix} = \begin{bmatrix}-W_{12}-W_{13}-W_{14} & W_{21} & W_{31} & W_{41}\\ W_{12} & -W_{21}-W_{23}-W_{24} & W_{32} & W_{42}\\ W_{13} & W_{23} & -W_{31}-W_{32}-W_{34} & W_{43}\\ W_{14} & W_{24} & W_{34} & -W_{41}-W_{42}-W_{43}\end{bmatrix}\begin{bmatrix}p_1\\p_2\\p_3\\p_4\end{bmatrix}, \tag{25}$$

where $W_{ij}$ and $W_{ji}$ are the spontaneous absorption and emission of phonons, respectively, between states $i$ and $j$ at temperature $T_0$. $W_{ij}$ can be estimated using thermodynamic relations $W_{ij} = W_0 n(E_{ij})$ and $W_{ji} =$ $W_0(n(E_{ij})+1)$, where $n(E_{ij})$ represents the occupation number of phonons with energy $E_{ij} = E_j - E_i$ and follows the Bose-Einstein distribution at temperature $T_0$. The negative signs behind the diagonal matrix elements are an indication of population reduction due to transition. The above set of rate equations is formulated and solved at times $t_1$and $t_4$ among states $|1\rangle, |2\rangle, |3\rangle$, and $|4\rangle$.

During the interval from $t_3$ to $t_4$, between states $|6\rangle$ and $|8\rangle$, there is coherent interaction with the microwave, causing Rabi oscillations. The Rabi oscillations are characterized by: $p_{46}(t) = \cos^2(\Omega t/2)$ and $p_8(t) = \sin^2(\Omega t/2)$, where Ω is the Rabi frequency. Their derivatives are: $dp_6(t)/dt = -(\Omega/2)\sin(\Omega t)$ and $dp_8(t)/dt = (\Omega/2)\sin(\Omega t)$. There are also spontaneous transitions among all these states, characterized by $W_{ij}$. Therefore, the population fraction rate equations in matrix form can be written as:

$$\frac{d}{dt}\begin{bmatrix} p_5 \\ p_6 \\ p_7 \\ p_8 \end{bmatrix} = \begin{bmatrix} -W_{56}-W_{57}-W_{58} & W_{65} & W_{75} & W_{85} \\ W_{56} & -W_{65}-W_{67}-W_{68} & W_{76} & W_{86} \\ W_{57} & W_{67} & -W_{75}-W_{76}-W_{78} & W_{87} \\ W_{58} & W_{68} & W_{78} & -W_{85}-W_{86}-W_{87} \end{bmatrix}\begin{bmatrix} p_5 \\ p_6 \\ p_7 \\ p_8 \end{bmatrix} + \begin{bmatrix} 0 \\ -(\Omega/2)\sin(\Omega t) \\ 0 \\ (\Omega/2)\sin(\Omega t) \end{bmatrix}. \quad (26)$$

The negative sign in the second term indicates a trade-off between the population of the states $|6\rangle$ and $|8\rangle$ due to Rabi oscillations, i.e., following an increase in the population of the state $|8\rangle$, a reduction in the population of the state $|6\rangle$ occurs.

### III-I. Cooling Cycle Process

The cooling cycle starts at $\epsilon(t_1)$, with state populations updated at each step, as outlined below:

- $\boldsymbol{t = t_1}$**, Initialization:** The population fractions in states $|1\rangle$ to $|4\rangle$ are initialized according to thermodynamic equilibrium at a temperature $T_0 = 1\,\mathrm{K}$, i.e., $p_i = \exp(-E_i/k_B T_0)/\sum_i \exp(-E_i/k_B T_0)$.
- $\boldsymbol{t_1}$ **to** $\boldsymbol{t_2^-}$**, Accelerated Adiabatic Transition from** $\boldsymbol{\epsilon(t_1)}$ **to** $\boldsymbol{\epsilon(t_2)}$**:** The populations of states $|1\rangle$ to $|4\rangle$ transition to states $|5\rangle$ to $|8\rangle$ with no entropy change. The work done on the system is given by $\Delta W_1 = \Delta U = \sum_{i=5,6,7,8} p_i(t_2)E_i(t_2) - \sum_{i=1,2,3,4} p_i(t_1)E_i(t_1)$.
- $\boldsymbol{t_2^+}$ **to** $\boldsymbol{t_3^-}$**, Coherent MW Interaction between States** $\boldsymbol{|6\rangle}$ **and** $\boldsymbol{|8\rangle}$**:** Rabi oscillations transfer the population from $|6\rangle$ to $|8\rangle$, while spontaneous phonon absorption moves the population from $|5\rangle$ to $|6\rangle$, and spontaneous phonon emission transfers the population from $|7\rangle$ to $|6\rangle$. As the oscillations proceed, states $|5\rangle$ and $|7\rangle$ gradually depopulate, reaching a small steady-state population, while the majority of the population oscillates between $|6\rangle$ and $|8\rangle$ (see Fig. 2). The MW is turned off once the population in state $|8\rangle$ reaches its maximum value. Minimal MW power is assumed so that it does not affect the system's energy states. The changes in energy and entropy are:
  - $\Delta U = \sum_{i=5,6,7,8} p_i(t_3^-)E_i(t_3^-) - \sum_{i=5,6,7,8} p_i(t_2^+)E_i(t_2^+)$,
  - $\Delta S = -k_B\left(\sum_{i=5,6,7,8} p_i(t_3^-)ln(p_i(t_3^-)) - \sum_{i=5,6,7,8} p_i(t_2^+)ln(p_i(t_2^+))\right)$.
- $\boldsymbol{t_3^+}$ **to** $\boldsymbol{t_4^-}$**, Fast Diabatic Transition from** $\boldsymbol{\epsilon(t_2)}$ **to** $\boldsymbol{\epsilon(t_1)}$**:** The majority of the population in $|8\rangle$ is transferred to states $|1\rangle$ to $|4\rangle$, with tunneling probabilities from $|8\rangle$ to $|3\rangle$ and $|8\rangle$ to $|2\rangle$. Spontaneous thermal absorption and emission between $|1\rangle$ and $|4\rangle$ also occur. A five-level system rate equation, extended form of eq. (25) to five levels, is solved to find the populations at $t_4^-$. For any remaining population in states $|5\rangle$ to $|7\rangle$, we assume transfer to $|2\rangle$ to $|4\rangle$, respectively.
- $\boldsymbol{t_4^+}$ **to** $\boldsymbol{t_4^+ + t_{\mathrm{dwell}}}$**, Phonon Absorption and Temperature Update:** The phonon system interacts continuously with the DQD. Given spin-lattice relaxation times in the order of tens of microseconds, we assume the DQD temperature remains unchanged during the cycle, updating it discretely at the end of each cycle. The four-level rate equation is solved to determine the populations of states $|1\rangle$ to $|4\rangle$. To account for the time when the DQD temperature was assumed constant during the back-and-forth transitions between $\epsilon(t_1)$ and $\epsilon(t_2)$, the system is allowed to thermalize for a period of $t_4 + t_{\mathrm{dwell}}$, rather than just $t_{\mathrm{dwell}}$, at this stage. The updated state populations at this stage serve as the initial values for the next cycle, starting at $t = t_4 + t_{\mathrm{dwell}}$.

### III-II. Numerical Results

For the numerical calculations, we assume the following parameters: Rabi frequency $\Omega = 2$ MHz, $g^* = 0.3$ (the effective $g$-factor of HHs in Ge), accelerated adiabatic transition time $t_{12} = 1$ μs, microwave radiation time $t_{23} =$ 5.25 μs, quantum dot separation $d = 50$ nm, the extent of the wave function $l = 25$ nm, effective mass $m^* = 0.36\, m_0$, $\mu_B = 2.58 \times 10^{-23}$J/T (corresponding to the density of states hole effective mass in Ge), detuning energy $\epsilon(t_2) = 55.8\ \mu\text{eV} - \text{U}$, U = 2 meV, and spin-conserving tunneling amplitude $t_N = 1$ μeV. The spin-flip tunneling matrix element $\lambda_2 = 0.1$ μeV, and the corresponding value between single-occupied states $|T_\pm(1,1)\rangle$ and $|S(1,1)\rangle$, $\lambda_1$ is 0.02 μeV, calculated as $\lambda_1 = \sqrt{2}\exp(-d^2/2l^2)\lambda_2$. $W_0$ is chosen as 50 kHz to fit the empirical value of the hole spin relaxation time $T_1$. $W_0 = 5$ kHz results in a hole spin relaxation time of 200 μ$s$ at 20 mK between the excited state $|T_-(1,1)\rangle$ and ground state $|S(0,2)\rangle$ at $\epsilon(t_1) + U = -65$ μeV. At this detuning energy, the energy splitting is 50 μeV corresponding to typical spin qubits. For consistency, $W_0$ is kept constant for all $W_{ij}$, and they only differ from each other through $n(E_{ij})$ term.

Figure 1b is plotted assuming these parameter values and $B = 0.3$ T. However, in the plots presented in this section, wherever the energy levels change with variables such as the magnetic field and tunneling matrix elements, they are calculated accordingly for consistency. The equivalent temperature $T$ at time $t = t_4$ is calculated by solving the following equation for $T$:

$$\frac{1}{Z(t_4)}\sum_{i=1,2,3,4} E_i(t_4)\exp(-\,E_i(t_4)/k_B T) = \sum_{i=1,2,3,4} p_i(t_4)E_i\,(t_4), \qquad (27)$$

where the partition function at $t = t_4$, is $Z(t_4) = \sum_{j=1,2,3,4}\exp(-E_j(t_4)/k_B T)$. This relation provides an effective temperature for which a four-level system in thermal equilibrium would have the same energy. However, in systems where spin populations across energy levels are not governed by a classical statistical distribution, defining a temperature in the conventional sense becomes more complex. In such non-equilibrium conditions, particularly when a system strongly interacts with a non-thermal driving force, temperature may not be well-defined in the traditional thermodynamic sense.

Figure 2 illustrates the transient population dynamics of states $p_5(t), p_6(t), p_7(t)$, and $p_8(t)$ starting at time $t_2$ with an initial temperature of 1 K during the first cycle of microwave radiation. The Rabi frequency is assumed to be 2 MHz. At the outset, the population in state $p_8(t_2)$ is much lower than the other states due to its higher energy. Microwave radiation, resonant with the energy difference between states $|6\rangle$ and $|8\rangle$, induces Rabi oscillations, causing the population to shift from $|6\rangle$ to $|8\rangle$. Concurrently, population transfer occurs from $|5\rangle$ to $|6\rangle$ via spontaneous absorption and from $|7\rangle$ to $|6\rangle$ via spontaneous emission. These interactions continuously pump the population from $|5\rangle$ and $|7\rangle$ into $|6\rangle$ and from $|6\rangle$ to $|8\rangle$. As Rabi oscillations proceed, states $|5\rangle$ and $|7\rangle$ become almost depopulated, and the population oscillates between $|6\rangle$ and $|8\rangle$ in resonance with the microwave radiation. This dynamic effectively drives the system toward the desired cooling state by manipulating the populations of the energy levels through controlled microwave interactions.

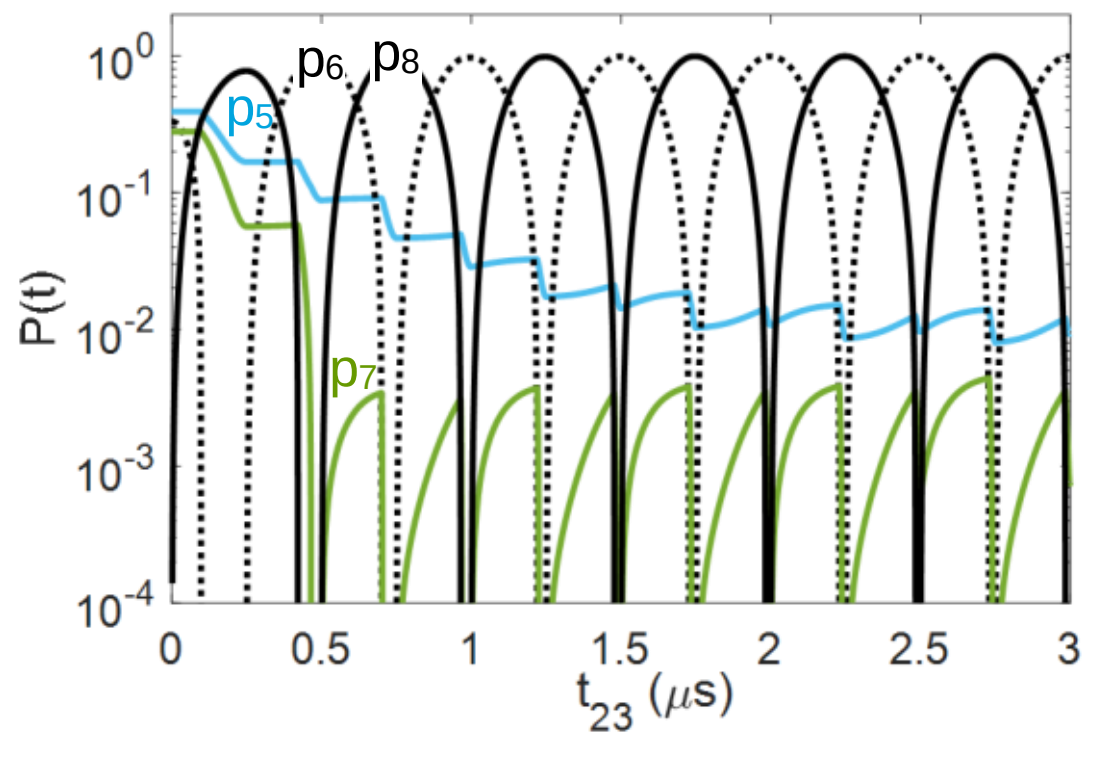

Figure 2: Transient population dynamics of states $p_5$ (solid blue), $p_6$ (dotted black), $p_7$ (solid green), and $p_8$ (solid black) starting at time $t_2$ with an initial temperature of 1 K during the first cycle of microwave radiation. The Rabi frequency is assumed to be 2 MHz. At the outset, the population in state $p_8(t_2)$) is much lower than in the other

states due to its higher energy. Microwave radiation, resonant with the energy difference between states |6⟩ and |8⟩, induces Rabi oscillations, causing the population oscillate between |6⟩ and |8⟩. Following a population shift from 6⟩ to |8⟩, population transfer occurs from |5⟩ to |6⟩ via spontaneous phonon absorption and from |7⟩ to |6⟩ via spontaneous phonon emission. These interactions continuously pump the population from |5⟩ and |7⟩ into |6⟩ and from |6⟩ to |8⟩. As the Rabi oscillations proceed, states |5⟩ and |7⟩ become almost depopulated, and the population oscillates between |6⟩ and |8⟩ in resonance with the microwave radiation.

Figure 3 provides an analysis of the system temperature and the normalized populations of the ground state ($p_1$) and first excited state ($p_2$) as functions of the detuning energy $\epsilon(t_1)+U$. Panel (a) illustrates the system temperature at $B=0.3$ T for three different bath temperatures: $T_0=1$ K (solid black line), $T_0=4$ K (solid blue line), and $T_0=10$ K (solid red line). At $\epsilon(t_1)+U\approx-20$ μeV and $t_{34}=10$ ns, the system temperature reaches 10 mK for $T_0=1$ K . As the detuning potential is further reduced, the temperature approaches 2 mK. When $t_{34}$ increases from 10 ns to 1 μs, the system temperature rises to 18 mK (the dotted black line). For bath temperatures of $T_0=4$ K and $T_0=10$ K with $t_{34}=10$ ns , the system temperature increases to 15 mK and 26 mK, respectively. Notably, at $T_0=10$ K and $\epsilon(t_1)+U\approx-15$ μeV, there exists an optimal detuning where the system temperature reaches a minimum, observed as a shoulder in $p_1/p_1^{eq}$ shown in panel (b). Panel (b) shows the normalized populations of the ground state ($p_1$ in black) and the first excited state ($p_2$ in blue) relative to their equilibrium values at steady state. As the system temperature decreases, the population of the ground state ($p_1$) increases, while the population of the excited state ($p_2$) decreases. This analysis emphasizes the sensitivity of the system temperature to the detuning energy and the diabatic return time, $t_{34}$, under varying bath temperatures.

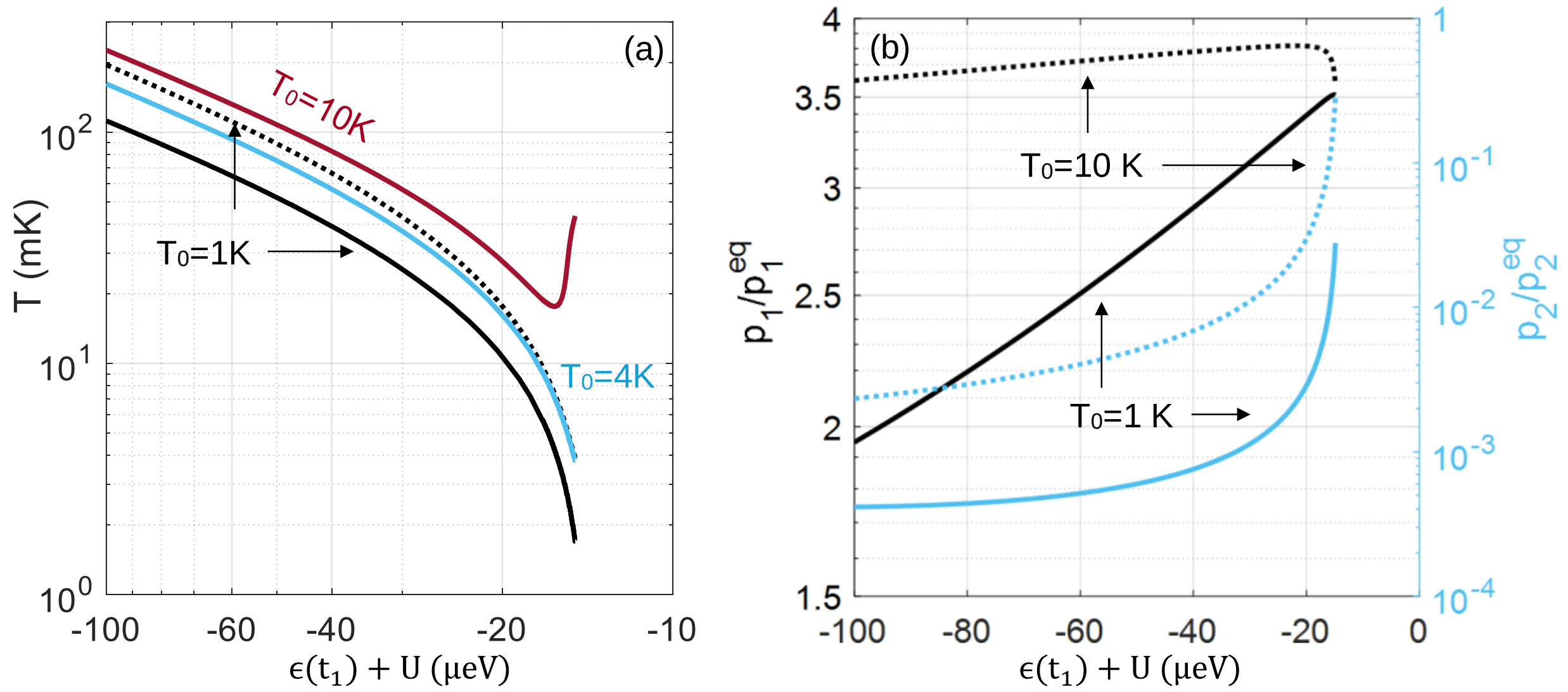

Figure 3: (a) System temperature and (b) normalized population of the ground state $p_1$ (blue) and the first excited state $p_2$ (black) at 1 K (solid line) and 10 K (dotted line) with respect to their equilibrium values at steady state versus detuning energy $\epsilon(t_1)+U$. In panel (a), the results are plotted for $B=0.3$ T at three bath temperatures: $T_0=1$ K (solid black), $T_0=4$ K (solid blue), and $T_0=10$ K (solid red). The system temperature at $\epsilon(t_1)+U\approx-20$ μeV and $t_{34}=10$ ns reaches 10 mK for $T_0=1$ K and approaches 2 mK as the detuning potential is further reduced. This temperature increases to 18 mK when $t_{34}$ rises from 10 ns to 1 μs (dotted black line). For $T_0=4$ K and 10 K, with $t_{34}=10$ ns, the system temperature increases to 15 mK and 26 mK, respectively. Panel (b) illustrates the effect of the detuning parameter on the populations of the ground state ($p_1$) and the first excited state ($p_2$). As detuning increases, both $p_1$ and $p_2$ populations generally rise, with $p_2$ exhibiting a steeper increase at higher detuning energies, and $p_1$ showing a turning point at the higher bath temperature (10 K), around $\epsilon(t_1)+U\approx-15$ μeV. As the system temperature increases, $p_1$ and $p_2$ increase accordingly.

Figure 4 provides a detailed examination of the system temperature's dependency on the fast diabatic return time ($t_{34}$) and the dwelling time ($t_{\mathrm{dwell}}$), which is the duration which the system interacts with phonons before initiating the next cooling cycle. These results highlight the critical role of the fast diabatic return time and the dwelling time in achieving and maintaining effective cooling in DQD systems. Both panels (a) and (b) assume a bath temperature ($T_0$) of 1 K and a detuning energy ($\epsilon(t_1)+U$) of approximately $-20$ μeV.

Panel (a) illustrates the system temperature as a function of the fast diabatic return time ($t_{34}$) for two different dwelling times: $t_{\mathrm{dwell}} = 10\ \mathrm{ns}$ (solid line) and $t_{\mathrm{dwell}} = 1\ \mu\mathrm{s}$ (dotted line). When $t_{\mathrm{dwell}} = 10\ \mathrm{ns}$, the system temperature remains between $10\ \mathrm{mK}$ and $20\ \mathrm{mK}$ over a wide range of $t_{34}$ from $1\ \mathrm{ns}$ to $20\ \mu\mathrm{s}$. However, for $t_{\mathrm{dwell}} = 1\ \mu\mathrm{s}$, the system temperature increases to a range of $40\ \mathrm{mK}$ to $50\ \mathrm{mK}$ over a similar range of $t_{34}$. This indicates that longer dwelling times can lead to higher system temperatures, highlighting the importance of optimizing $t_{\mathrm{dwell}}$ for effective cooling. Panel (b) depicts the system temperature as a function of the dwelling time for two different fast diabatic return times: $t_{34} = 10\ \mathrm{ns}$ (solid line) and $t_{34} = 1\ \mu\mathrm{s}$ (dotted line). The temperature increases with $t_{\mathrm{dwell}}$, as expected, and saturates at approximately $1\ \mathrm{K}$ (the bath temperature) when $t_{\mathrm{dwell}}$ approaches $100\ \mu\mathrm{s}$. In practice, the dwelling time should be optimized to be short enough to maintain a low system temperature, yet long enough to allow the spins to interact and effectively filter phonons through absorption.

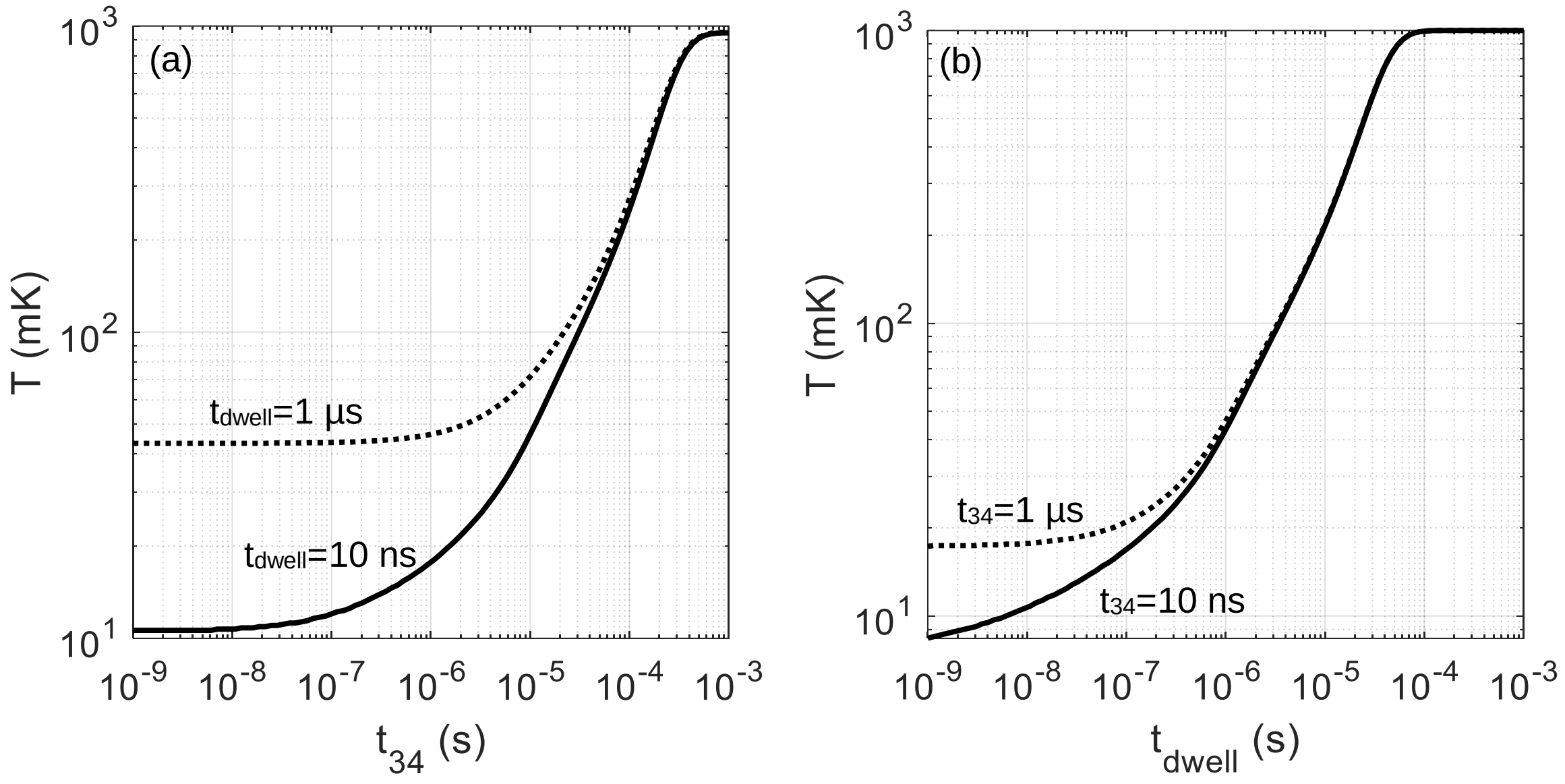

Figure 4: (a) System temperature versus fast diabatic return time $t_{34}$ and (b) system temperature versus dwelling time $t_{\mathrm{dwell}}$, which is the duration when the system interacts with phonons before starting the next cooling cycle. (a) shows the results for $t_{\mathrm{dwell}} = 10\ \mathrm{ns}$ (solid line) and $t_{\mathrm{dwell}} = 1\ \mu\mathrm{s}$ (dotted line). The system temperature for $t_{\mathrm{dwell}} = 10\ \mathrm{ns}$ ranges between $10\ \mathrm{mK}$ and $20\ \mathrm{mK}$ over a wide range of fast diabatic return times from $1\ \mathrm{ns}$ to $20\ \mu\mathrm{s}$. For $t_{\mathrm{dwell}} = 1\ \mu\mathrm{s}$, the system temperature increases to between $40\ \mathrm{mK}$ and $50\ \mathrm{mK}$ over a similar range of $t_{34}$. (b) shows the results for $t_{34} = 10\ \mathrm{ns}$ (solid line) and $t_{34} = 1\ \mu\mathrm{s}$ (dotted line). The temperature increases with $t_{\mathrm{dwell}}$ as expected, saturating at approximately $1\ \mathrm{K}$ when $t_{\mathrm{dwell}} \approx 100\ \mu\mathrm{s}$. Both panels (a) and (b) assume $T_0 = 1\ \mathrm{K}$ and $\epsilon(t_1) + U \approx -20\ \mu\mathrm{eV}$.

### III-III. Clarifying Energy Conservation During Gate Voltage Transitions

A central issue in this system is clarifying the mechanism of energy conservation during gate voltage transitions. One might initially assume that these transitions release energy as photons or phonons. However, when transitioning from $|S(0,2)\rangle$ at $\epsilon(t_2)$ to $|S(0,2)\rangle$ at $\epsilon(t_1)$, no charge transfer occurs between the quantum dots, ruling out photon or phonon emission due to charge movement.

Instead, this process is not isentropic, resulting in heat exchange between the double quantum dot (DQD) and the environment. Consequently, energy is either emitted or absorbed in the form of phonons, depending on the sign of $\Delta S = S(t_1) - S(t_2)$. However, due to the lack of charge movement between the QDs, no work is done on the system.

The accurate explanation for energy conservation here is that energy is conserved by returning it to the power supply. The gate and quantum dot system can be modeled as a capacitor, where the power supply provides energy to increase the electrostatic energy of the holes as the gate voltage shifts the system from $\epsilon(t_1)$ to $\epsilon(t_2)$. This energy increase is especially significant for the $|S(0,2)\rangle$ state, as shown in Figure 1b. When the gate voltage switches back to a negative value, the energy is returned to the power supply.

Thus, this transition has two concurrent effects: an increase in entropy ($\Delta S > 0$) through heat absorption from the lattice, and a reduction in internal energy ($\Delta U < 0$) as energy is returned to the power supply.

Additionally, microwave absorption during time $t_{23}$ adds energy to the system by increasing the population fraction in the $|S(0,2)\rangle$ state. Consequently, when the voltage switches back to negative, more energy is returned to the power supply than was initially provided, resulting in a lower total energy of the system.

Furthermore, one might consider that changes in the wave function distribution in the DQD, caused by cyclic detuning, could lead to photon emission. However, this effect is contingent on the rate of change in the gate voltage. Since $\Delta U$ is determined by the fixed energy states $E_i$ and the final populations $p_i$ at times $t_1$ and $t_4$, the released energy remains independent of the cyclic detuning rate. Therefore, photon emission due to detuning rate cannot account for the energy conservation.

### IV. Why Do We Need a DQD System?

A natural question arises: why not achieve excited state depopulation in a two-level system using a single QD with coherent MW interaction? While Rabi oscillations can indeed alternately populate and depopulate the ground and excited states in a single QD, and one could turn off the MW field at the moment the excited state is fully depopulated, this approach has drawbacks. In this scenario, energy is only conserved until the MW field is turned off. At that point, if the excited state is depopulated (as indicated by the pink point in Figure 6), a photon is emitted, which will likely be absorbed by the surrounding material, potentially causing unwanted lattice heating.

With a DQD system, however, we create a state inversion where the ground state $|S(0,2)\rangle$ at $\epsilon(t_1)$ becomes an excited state $|S(0,2)\rangle$ at $\epsilon(t_2)$. In this configuration, when the MW is turned off after populating this state, the photon is absorbed by the spin rather than being emitted into the lattice (similar to the blue point in Figure 6). The system is then returned to $\epsilon(t_1)$ through electrostatic potential adjustments, allowing the excess energy to be returned to the power supply. This process minimizes MW heating or eliminates it in the ideal case.

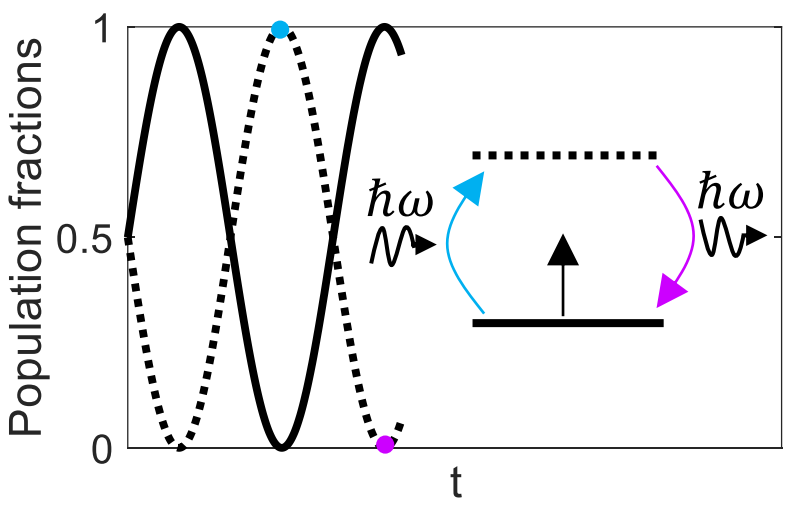

Figure 5: Schematic comparison of single QD versus DQD systems during MW interaction. In the single QD setup (pink point), emitted photons may lead to lattice heating, while in the DQD system (blue point), state inversion allows photon absorption, minimizing or eliminating MW heating.

### V. Practical Considerations

For practical implementation of the accelerated adiabatic transition, the ideal pulse shape derived from the protocol can be discretized into a series of linear ramps to mimic experimentally feasible pulses.

The second consideration is for microwave excitation at time $t_2$. At a detuning of $\epsilon(t_2) + U = 55.8\ \mu\text{eV}$ and $B = 0.3$ T, the energy difference between $|S(1,1)\rangle$ and $|S(0,2)\rangle$ is $41\ \mu\text{eV}$, requiring a microwave excitation frequency of $10$ GHz for the $|S(1,1)\rangle$ to $|S(0,2)\rangle$ transition. Increasing the detuning energy at time $t_2$ results in a higher energy difference. Therefore, the system can be designed by adjusting $\epsilon(t_2)$ to match the available microwave sources in the laboratory. Other parameters, such as the magnetic field and geometrical dimensions of the dots, would also affect the required microwave frequency. These adjustments are crucial for ensuring the efficient operation of the cooling loop within the experimental setup.

Another practical consideration is the sensitivity of the energy levels to the DQD parameters. Key parameters include $\lambda_1$, $\lambda_2$, $t_N$, $g^*$, and $\mu_B$. The energy levels are significantly more sensitive to the values $g^*$, $t_N$, and $\mu_B$ than $\lambda_1$ and $\lambda_2$ . The dots' separation $d$ is usually a multiple of the extent of the wavefunction $l$. Considering $d = 2l$, the overlap parameter is $S\sim 0.14$, and $\lambda_1/\lambda_2 \sim 0.2$. This indicates that the transition between different charge configurations, $|T_{\pm}(1,1)\rangle$ and $|S(0,2)\rangle$, due to SOC is more probable compared to the transition between $|T_{\pm}(1,1)\rangle$ and $|S(1,1)\rangle$. For materials with strong SOC, like GaAs, this ratio is about $10^{-2}$, and without loss of generality, one can ignore the effects of the spin-flip tunneling matrix element $\lambda_1$.

As an illustration, Figure 6 demonstrates how the system temperature varies with $\lambda_2$ for two different values of $\lambda_1$: $0.015\lambda_2$ and $0.2\lambda_2$. The system temperature remains relatively stable across a wide range of $\lambda_2$. However, the temperature increases significantly when $\lambda_2$ exceeds the spin-conserved tunneling energy $t_N$, which can occur under very large SOC.

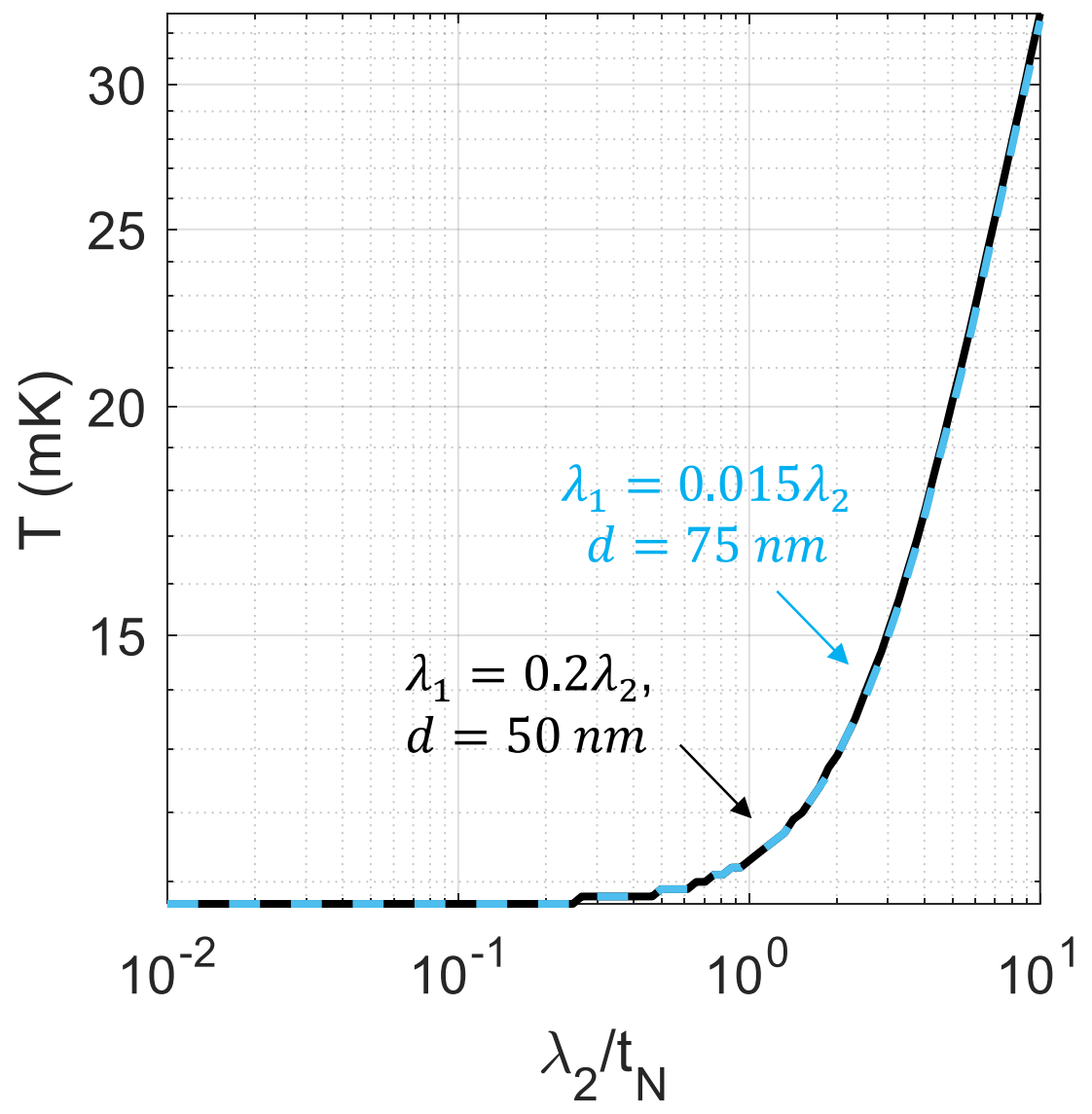

Figure 6: System temperature as a function of $\lambda_2$ for two different values of $\lambda_1 = 0.2\lambda_2$ and $0.015\lambda_2$. Assuming the extension of the wavefunction $l$ is 25 nm, the corresponding distances between the QDs are $d = 50$ nm and 75 nm. The system temperature remains relatively stable over a wide range of $\lambda_2$. However, at very high values of $\lambda_2$, where spin-flip tunneling exceeds spin-conserved tunneling, the system temperature increases with $\lambda_2$. The system temperature shows little sensitivity to changes in $\lambda_1$.

Another practical consideration is the importance of the accelerated adiabatic transition in achieving the predicted low temperatures. Given that the energy states can be very close to each other depending on the detuning energy, achieving complete adiabatic transitions might be challenging in practice. Figure 7 illustrates the system temperature as a function of detuning energy at $t_1$, considering the population transfer from states $|1\rangle$, $|2\rangle$, $|3\rangle$, and $|4\rangle$ to states $|5\rangle$, $|6\rangle$, $|7\rangle$, and $|8\rangle$ via two different paths: accelerated adiabatic transfer and complete population transfer to $|7\rangle$ ($|T_{+}(1,1)\rangle$). The path where the population is transferred to $|7\rangle$ results in a higher system temperature; however, this increase is not significant. Consequently, while accelerated adiabatic transitions enhance performance, they are not strictly necessary for the cooling loop to function effectively and can be relaxed without a substantial loss in the achievable low temperature.

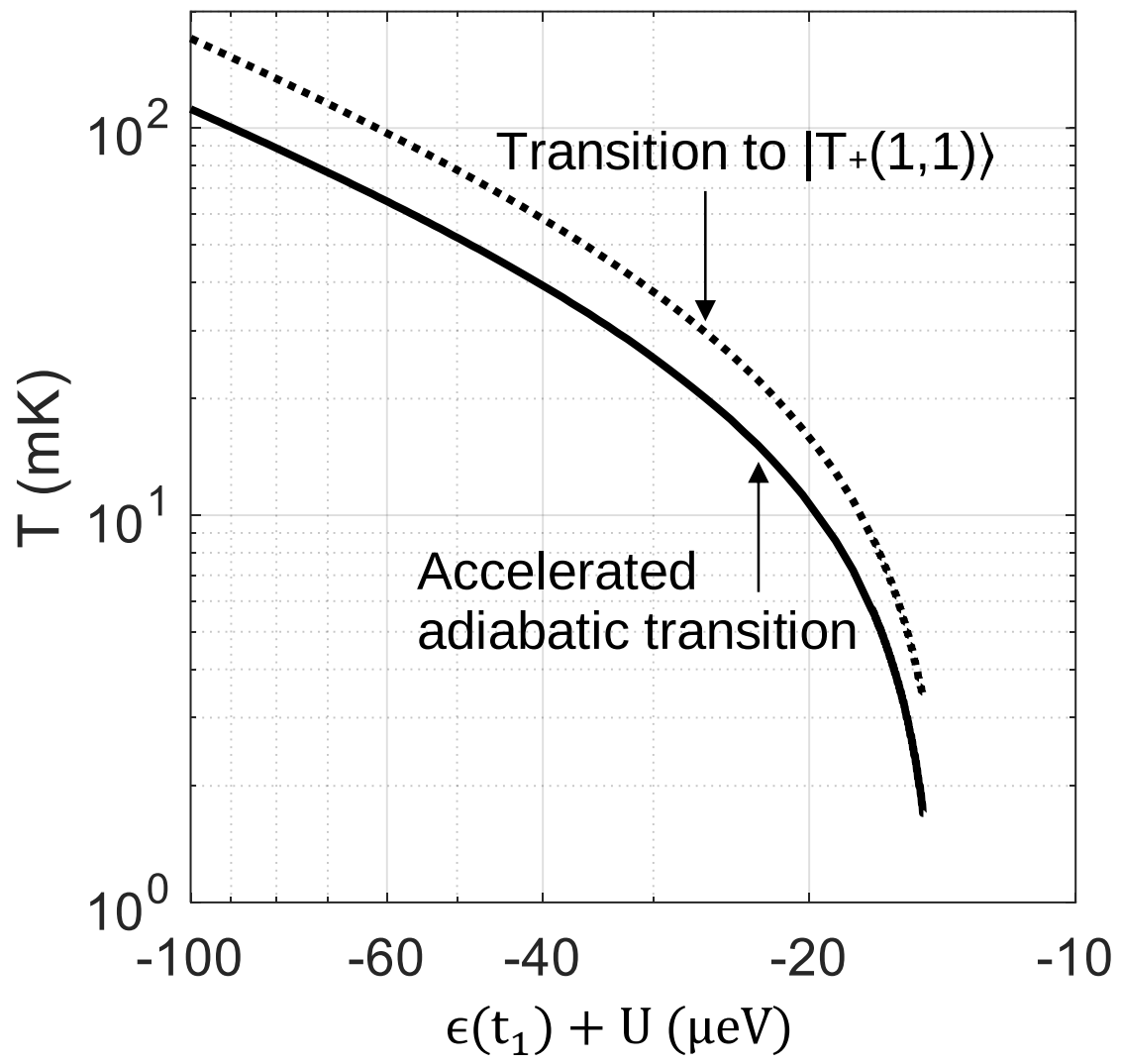

Figure 7: System temperature versus detuning energy at $t_1$ assuming the population is transferred from states $|1\rangle$, $|2\rangle$, $|3\rangle$, and $|4\rangle$ to $|5\rangle$, $|6\rangle$, $|7\rangle$, and $|8\rangle$ via two different paths: accelerated adiabatic transfer, and all population transfer to $|7\rangle$. The second path, where all the population is transferred to $|7\rangle$ ($|T_+(1,1)\rangle$), results in a higher system temperature; however, the difference is not significant. Therefore, the accelerated adiabatic transition is not strictly necessary for the cooling loop to function and can be relaxed without a substantial loss in the achievable low temperature.

This behavior occurs because, at time $t_3$, regardless of the initial population distribution among states $|5\rangle$, $|6\rangle$, $|7\rangle$, and $|8\rangle$, microwave radiation resonant with the energy difference between states $|6\rangle$ and $|8\rangle$ induces Rabi oscillations, causing the population to shift from $|6\rangle$ to $|8\rangle$. Concurrently, population transfer occurs from $|5\rangle$ to $|6\rangle$ via spontaneous absorption and from $|7\rangle$ to $|6\rangle$ via spontaneous emission. These interactions continuously pump the population from $|5\rangle$ and $|7\rangle$ into $|6\rangle$ and from $|6\rangle$ to $|8\rangle$. As the Rabi oscillations proceed, states $|5\rangle$ and $|7\rangle$ become almost depopulated, and the population oscillates between $|6\rangle$ and $|8\rangle$ in resonance with the microwave radiation. This dynamic effectively drives the population to $|8\rangle$ regardless of the initial population distribution, explaining why the different paths result in similar final system temperatures.

The lowest achievable temperature may vary with different values of magnetic field ($B$), as the cooling effect arises from the interplay of relative energy level spacings that influence the thermal excitation rate. Figure 8 presents a contour map of the system temperature as a function of the magnetic field and detuning energy at time $t_1$. A minimum temperature of $2\,\mathrm{mK}$ is achievable under optimal parameters. Each temperature corresponds to a specific range of magnetic field and detuning energy ($\epsilon(t_1) + U$). For instance, to achieve a temperature of $2\ \mathrm{mK}$, the detuning energy can range from near zero to $-50\ \mu\mathrm{eV}$, with the magnetic field between near zero and $1\ \mathrm{T}$. As the temperature increases, these constraints become more relaxed, providing greater flexibility for experimental design.

It is important to note that some points in the contour map may not correspond to the pure states shown in Figure 1 at time $t_1$. Therefore, the optimal detuning energy and magnetic field must be determined based on the specific parameters of the DQD system. The lowest achievable temperature is ultimately determined by the competition between state population changes due to thermal excitation and microwave-induced state depopulation. Consequently, in practice, the DQD parameters must be carefully considered, and the energy levels must be measured or accurately calculated when designing the cooling loop.

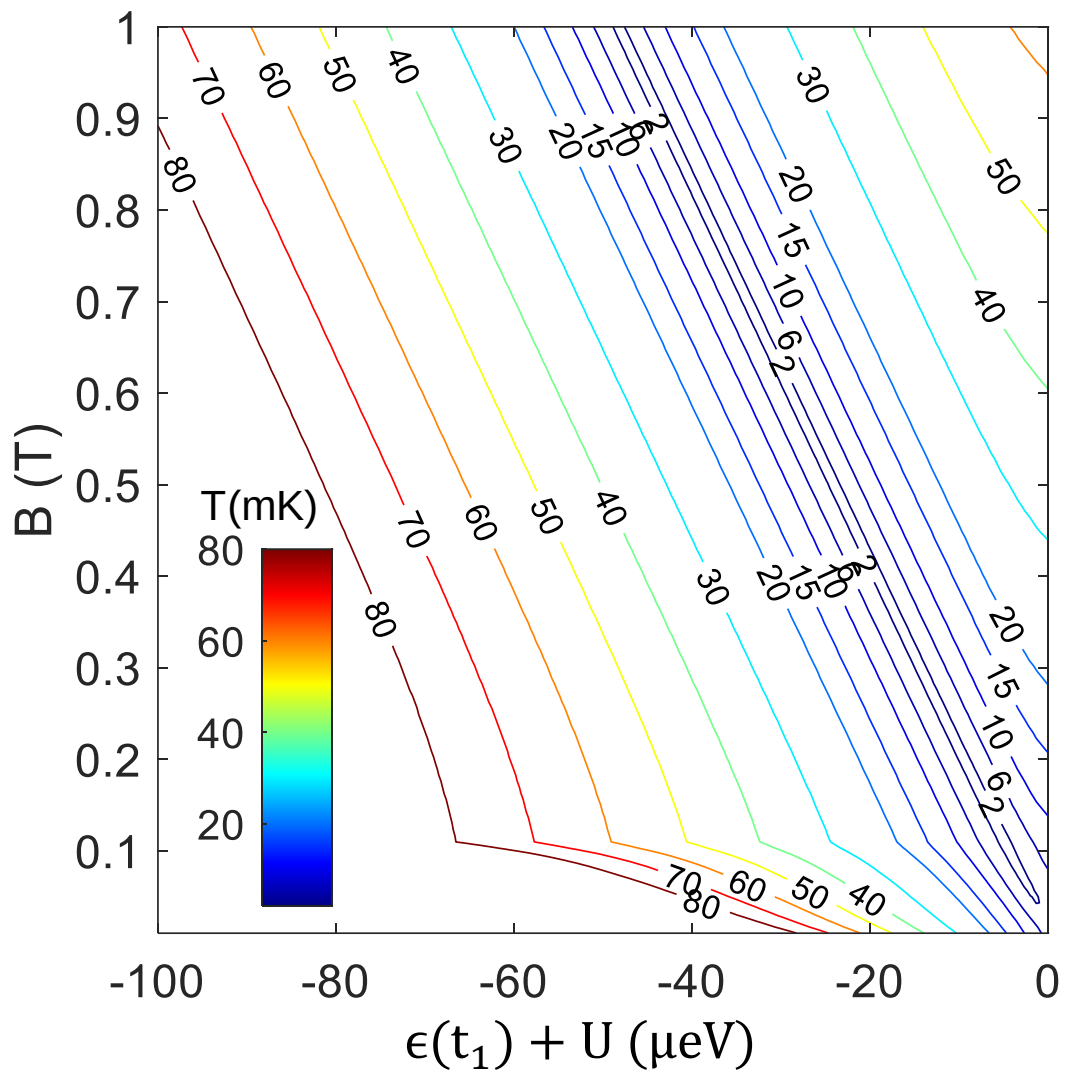

Figure 8: Contour map of the system temperature as a function of the magnetic field and detuning energy at time $t_1$. A minimum temperature of $2\ \mathrm{mK}$ is achievable under optimal parameters. Each temperature corresponds to a specific range of magnetic field and detuning energy ($\epsilon(t_1)+U$). For instance, to achieve a temperature of $2\ \mathrm{mK}$, the detuning energy can range from near zero to $-50\ \mu\mathrm{eV}$, with the magnetic field between near zero and $1\ \mathrm{T}$.

Additionally, the dwelling time ($t_{\mathrm{dwell}}$) between cycles is a crucial parameter. This time must be optimized so that the DQD has sufficient time to absorb specific phonons, yet remains short enough to prevent the local environment around the DQD from becoming isothermal with the rest of the lattice. The appropriate duration depends on factors such as the material properties, operating temperature, and the area intended for local cooling to effectively manage the nearby computational spin qubits.

The primary role of the DQD cooler discussed in this paper is not to cool the lattice surrounding computational qubits, but to selectively filter out phonons that are in resonance with the qubits. By absorbing these resonant phonons, the DQD drives the phonon population into a non-equilibrium distribution across different phonon frequencies, meaning that assigning a single temperature to the system does not adequately describe its thermal state. The critical point is that the DQD filters out the phonons most responsible for qubit decoherence, which is expected to increase the coherence time of computational qubits in its vicinity. As such, traditional thermodynamic calculations like the coefficient of performance (COP) are not relevant for this purpose. Instead, the key metric that should be calculated is the decoherence time. However, this paper does not explore the effect on qubit coherence directly, this requires a dedicated investigation, which will be the focus of our future studies.

Another important consideration is the hole spin phase shift caused by interactions with phonons, particularly during the thermal transition from states $|6\rangle$ to $|8\rangle$. In this transition, a hole tunnels between the left and right dots, and interacts with a phonon field via the following Hamiltonian:

$$H_{h-ph} = \sum_{\vec{q}}[(\alpha_{\vec{q}}|S(1,1)\rangle\langle S(1,1)| + \beta_{\vec{q}}|S(0,2)\rangle\langle S(0,2)|)(a_{-\vec{q}} + a_{\vec{q}}^{\dagger})] \quad (28)$$

where $a_{\vec{q}}^{\dagger}$ is creation operator for a phonon of mode $\vec{q}$, and $\alpha_{\vec{q}}$ and $\beta_{\vec{q}}$ are hole-phonon matrix elements. In general, the hole-phonon coupling can be asymmetric ($\alpha_{\vec{q}} \neq \beta_{\vec{q}}$). The hole-phonon interaction locally changes the energy of the hole, depending on whether the hole is in the left dot or in the right. During tunneling, hole's wave function experiences an additional phase shift due to this coupling. This phase shift is zero if the coupling is identical in both dots ($\alpha_{\vec{q}} = \beta_{\vec{q}}$). In the case of non-identical coupling, the phase shift depends on the state of the phonon field ($e^{i\vec{q}.\vec{d}}$ with $|\vec{d}| = d$, the separation of QDs). In our model calculations this effect can be practically compensated by adjusting the microwave pulse width to ensure p6 is minimized and p8 is maximized at the end of the microwave pulse.

In relation to cooling systems generally, quantum correlations and entanglement in the eigenstates of the Hamiltonian can negatively impact the achievement of desired temperatures. For instance, in spin-star cooling systems, the Heisenberg interaction between a central spin and surrounding ancilla qubits includes a transverse component, leading to entanglement between the central and ancilla qubits. This entanglement generation is detrimental to the cooling process.[9] Conversely, when qubit interactions follow the Ising model, the ground state and all excited states of the spin-star system are factorized, resulting in negligible entanglement. Consequently, the minimum achievable temperature in the Ising model is lower than that in the Heisenberg model.

Therefore, it is pertinent to examine entanglement behavior in the context of our DQD cooling system. Quantum Fisher Information (QFI), a measure of multipartite entanglement,[15] has recently been studied for QDs with a universal Hamiltonian.[16] As demonstrated in [16], exchange interactions and confinement effects can induce unexpected entanglement within the system, which may influence the cooling mechanism. Understanding these entanglement dynamics will be valuable for refining our cooling approach and enhancing its effectiveness.

## VI. Conclusion

This paper presents a novel cooling mechanism that uses gate-defined double quantum dots (DQDs) as phonon filters, rather than conventional coolers, to reduce local phonon noise around computational qubits. By absorbing phonons resonant with qubit energy levels, the DQDs create a non-equilibrium phonon distribution, effectively enhancing qubit coherence. The cooling process consists of two key steps. First, microwave-induced state depopulation, combined with fast cyclic detuning, transfers the DQD's population to the ground state, lowering its temperature. Second, the cooled DQD filters out phonons that contribute to spin-lattice relaxation, thereby reducing decoherence in nearby qubits.

Numerical calculations demonstrate the feasibility of achieving local DQD temperatures below $10\ \mathrm{mK}$ at a bath temperature of $1\ \mathrm{K}$. The cooling performance is highly sensitive to parameters such as detuning energy, magnetic field strength, and diabatic return time. While this study primarily focuses on the first step – state depopulation and temperature reduction – the second step, involving phonon filtering and its impact on qubit coherence, requires further investigation. This approach addresses the critical challenge of suppressing decoherence in multi-qubit systems by minimizing thermal noise at the qubit level.

**Acknowledgement**

This study is partially based on work supported by AFOSR and LPS under contract numbers FA9550-23-1-0302 and FA9550-23-1-0763, and by the NSF under grant number CBET-2110603.

## References

[1] Loss, D., & DiVincenzo, D. P. (1998). Quantum computation with quantum dots. Physical Review A, 57(1), 120.
[2] Vandersypen, L. M. K., et al. (2017). Interfacing spin qubits in quantum dots and donors—hot, dense, and coherent. npj Quantum Information, 3(1), 1-10.
[3] Petit, L., et al. (2020). Universal quantum logic in hot silicon qubits. Nature, 580(7803), 355-359.
[4] Larson, D. J., et al. (1986). Sympathetic cooling of trapped ions: A laser-cooled two-species nonneutral ion plasma. Physical Review Letters, 57(1), 70
[5] Bohman, M., et al. (2018). Sympathetic cooling of protons and antiprotons with a common endcap Penning trap. Journal of Modern Optics, 65(5-6), 568-576
[6] Grajcar, M., et al. (2008). Sisyphus cooling and amplification by a superconducting qubit. Nature Physics, 4(8), 612-616
[7] Boykin, P. O., et al. (2002). Algorithmic cooling and scalable NMR quantum computers. Proceedings of the National Academy of Sciences, 99(6), 3388-3393.
[8] Fernandez, J. M., et al. (2004). Algorithmic cooling of spins: A practicable method for increasing polarization. International Journal of Quantum Information, 2(04), 461-477
[9] Arısoy, O., & Müstecaplıoğlu, Ö. E. (2021). Few-qubit quantum refrigerator for cooling a multi-qubit system. Scientific Reports, 11(1), 12981
[10] Mitchison, M. T., Woods, M. P., Prior, J., & Huber, M. (2015). Coherence-assisted single-shot cooling by quantum absorption refrigerators. *New Journal of Physics*, 17(11), 115013
[11] Brask, Jonatan Bohr, and Nicolas Brunner. (2015). "Small quantum absorption refrigerator in the transient regime: Time scales, enhanced cooling, and entanglement." Physical Review E 92, no. 6, 062101.
[12] Hewgill, Adam, J. Onam González, José P. Palao, Daniel Alonso, Alessandro Ferraro, and Gabriele De Chiara. (2020). "Three-qubit refrigerator with two-body interactions." Physical Review E 101, no. 1, 012109.
[13] Arısoy, Onat, Steve Campbell, and Özgür E. Müstecaplıoğlu. (2019). "Thermalization of finite many-body systems by a collision model." Entropy 21, no. 12, 1182.
[14] Fernández-Fernández, D., Yue Ban, and G. Platero. (2022). "Quantum control of hole spin qubits in double quantum dots." *Physical Review Applied* 18, no. 5, 054090.
[15] P. Hauke, M. Heyl, L. Tagliacozzo, and P. Zoller. (2016). "Measuring multipartite entanglement through dynamic susceptibilities" Nat. Phys. 12, 778.
[16] J. Abouie, and D. Vashaee. (2024). "Entanglement in Quantum Dots: Insights from Dynamic Susceptibility and Quantum Fisher Information" Advanced Quantum Technology 2024, 2400117.